\documentclass[titlepage,12pt]{article}
\usepackage[dvips]{graphicx}
\begin{document}

\title{Confinement of spin-0 and spin-1/2 particles\\
in a mixed vector-scalar coupling\\
with unequal shapes for the potentials}
\date{}
\author{Luis B. Castro\thanks{%
E-mail address: benito@feg.unesp.br (L.B. Castro)} and Antonio S. de Castro%
\thanks{%
E-mail address: castro@pesquisador.cnpq.br (A.S. de Castro)} \\
\\
UNESP - Campus de Guaratinguet\'{a}\\
Departamento de F\'{\i}sica e Qu\'{\i}mica\\
12516-410 Guaratinguet\'{a} SP - Brasil }
\date{}
\maketitle

\begin{abstract}
The Klein-Gordon and the Dirac equations with vector and scalar potentials
are investigated under a more general condition, $V_{v}=V_{s} + \mathrm{%
const.}$ These isospectral problems are solved in a case of squared
trigonometric potential functions and bound states for either particles or
antiparticles are found. The eigenvalues and eigenfuntions are discussed in
some detail. It is revealed that a spin-0 particle is better localized than
a spin-1/2 particle when they have the same mass and are subject to the same
potentials.
\end{abstract}

There has been a continuos interest for solving the Klein-Gordon (KG) and
the Dirac equations in the four-dimensional space-time as well as in lower
dimensions for a variety of potentials. It is well known from the quarkonium
phenomenology that the best fit for meson spectroscopy is found for a
convenient mixture of vector and scalar potentials put by hand in the
equations (see, e.g., \cite{luc}). The same can be said about the treatment
of the nuclear phenomena describing the influence of the nuclear medium on
the nucleons \cite{ser}. The mixed vector-scalar potential has also been
analyzed in 1+1 dimensions. In this mixed two-dimensional context, all the
works has been devoted to the investigation of the solutions of the
relativistic equations by assuming that the vector and scalar potential
functions are proportional \cite{gum}. In the present work the problem of
relativistic particles is considered with a mixing of vector and scalar
Lorentz structures with unequal potential functions. The mixing for this
enlarged class of problems is chosen in such a way that the difference
between the vector and the scalar potential functions is a constant. Except
for a possible isolated solution for the Dirac equation, the KG equation and
the Dirac equation for the upper component of the Dirac spinor are both
mapped into a Schr\"{o}dinger-like equation. Squared trigonometric potential
functions are chosen in such a way that these relativistic problems are
mapped into a Sturm-Liouville problem with the exactly solvable effective
symmetric P\"{o}schl-Teller potential \cite{rm}-\cite{nie1}. Then, the whole
relativistic spectrum is found, if the particle is massless or not. The
process of solving the KG and the Dirac equations for the eigenenergies has
been transmuted into the simpler and more efficient process of solving an
irrational algebraic equation. Apart from the intrinsic interest as new
solutions of fundamental equations in physics, the bound-state solutions of
these systems are important in condensed matter mainly because of their
potential applications ranging from ferroelectric domain walls in solids and
magnetic chains \cite{bra}.

In the presence of vector and scalar potentials the 1+1 dimensional
time-inde\-pen\-dent KG equation for a particle of rest mass $m$ reads

\begin{equation}
-\hbar ^{2}c^{2}\,\phi ^{\prime \prime }+\left( mc^{2}+V_{s}\right) ^{2}\phi
=\left( E-V_{v}\right) ^{2}\phi  \label{1}
\end{equation}

\noindent where the prime denotes differentiation with respect to $x$, $E$
is the energy of the particle, $c$ is the velocity of light and $\hbar $ is
the Planck constant. The subscripts for the terms of potential denote their
properties under a Lorentz transformation: $v$ for the time component of the
2-vector potential and $s$ for the scalar term. In the presence of
time-in\-de\-pen\-dent vector and scalar potentials the 1+1 dimensional
time-in\-de\-pen\-dent Dirac equation for a fermion of rest mass $m$ reads

\begin{equation}
\left[ c\alpha p+\beta \left( mc^{2}+V_{s}\right) +V_{v}\right] \psi =E\psi
\label{4}
\end{equation}

\noindent where $p$ is the momentum operator. $\alpha $ and $\beta $ are
Hermitian square matrices satisfying the relations $\alpha ^{2}=\beta ^{2}=1$%
, $\left\{ \alpha ,\beta \right\} =0$. From the last two relations it
follows that both $\alpha $ and $\beta $ are traceless and have eigenvalues
equal to $\pm $1, so that one can conclude that $\alpha $ and $\beta $ are
even-dimensional matrices. One can choose the 2$\times $2 Pauli matrices
satisfying the same algebra as $\alpha $ and $\beta $, resulting in a
2-component spinor $\psi $. We use $\alpha =\sigma _{1}$ and $\beta =\sigma
_{3}$. Provided that the spinor is written in terms of the upper and the
lower components, $\psi _{+}$ and $\psi _{-}$ respectively, \noindent the
Dirac equation decomposes into:
\begin{equation}
i\hbar c\psi _{\pm }^{\prime }=\left[ V_{v}-E\mp \left( mc^{2}+V_{s}\right) %
\right] \psi _{\mp }  \label{5}
\end{equation}%
In the nonrelativistic approximation (potential energies small compared to $%
mc^{2}$ and $E\simeq mc^{2}$) Eq. (\ref{1}) becomes the Schr\"{o}dinger
equation with binding energy equal to $E-mc^{2}$ and a potential given by $%
V_{v}+V_{s}$, \noindent so that $\phi $ obeys the Schr\"{o}dinger equation
without distinguishing the contributions of vector and scalar potentials. In
this approximation Eq. (\ref{5}) becomes $\psi _{-}=p/(2mc)\psi _{+}$,
\noindent and because of this $\psi _{+}$ obeys the same equations as $\phi $
\ while $\psi _{-}$ is of order $v/c<<1$ relative to $\psi _{+}$. It is
remarkable that the KG and the Dirac equations with a scalar potential, or a
vector potential contaminated with some scalar coupling, is not invariant
under the simultaneous changes $V\rightarrow V+\mathrm{const.}$ and $%
E\rightarrow E+\mathrm{const.}$, this is so because only the vector
potential couples to the charge, whereas the scalar potential couples to the
mass of the particle. Therefore, if there is any scalar coupling the energy
itself has physical significance and not just the energy difference.

It is well known that a confining potential in the nonrelativistic approach
is not confining in the relativistic approach when it is considered as a
Lorentz vector. It is surprising that relativistic confining potentials may
result in nonconfinement in the nonrelativistic approach, simply because
there is pair creation and the single-particle picture no long holds. This
last phenomenon is a consequence of the fact that vector and scalar
potentials couple differently in the KG and in the Dirac equations whereas
there is no such distinction among them in the Schr\"{o}dinger equation.
This observation permit us to conclude that even a \textquotedblleft
repulsive\textquotedblright\ potential can be a confining potential. The
case $V_{v}=-V_{s}$ presents bounded solutions in the relativistic approach,
although it reduces to the free-particle problem in the nonrelativistic
limit. The attractive vector potential for a particle is, of course,
repulsive for its corresponding antiparticle, and vice versa. However, the
attractive (repulsive) scalar potential for particles is also attractive
(repulsive) for antiparticles. For $V_{v}=V_{s}$ and an attractive vector
potential for particles, the scalar potential is counterbalanced by the
vector potential for antiparticles as long as the scalar potential is
attractive and the vector potential is repulsive. As a consequence there is
no bounded solution for antiparticles. For $V_{v}=0$ and a pure scalar
attractive potential, one finds energy levels for particles and
antiparticles arranged symmetrically about $E=0$. For $V_{v}=-V_{s}$ and a
repulsive vector potential for particles, the scalar and the vector
potentials are attractive for antiparticles but their effects are
counterbalanced for particles. Thus, recurring to this simple standpoint one
can anticipate in the mind that there is no bound-state solution for
particles in this last case of mixing. Regarding the structure of the
wavefunctions under the simultaneous changes $V_{v}\rightarrow -V_{v}$ and $%
E\rightarrow -E$, from the charge-conjugation operation one can see that if $%
\psi $ is a solution with energy $E$ for the potential $V_{v}$ then $\sigma
_{1}\psi ^{\ast }$ is also a solution with energy $-E$ for the potential $%
-V_{v}$. Thus, one has $\left( \psi _{\pm }\right) _{c}=\psi _{\mp }^{\ast }$
and that means that the upper and lower components of the Dirac spinor have
their roles changed. As for the KG wavefunction, its nodal structure is
trivially preserved in such a way that particle and antiparticle can be
distinguished only by the eigenenergies.

Supposing that the vector and scalar potentials are constrained by the
relation $V_{v}-V_{s}=V_{0}$, where $V_{0}$ is a constant, and defining

\begin{equation}
\varepsilon =E-V_{0},\qquad E_{\mathtt{eff}}=\frac{\varepsilon
^{2}-m^{2}c^{4}}{2mc^{2}},\qquad V_{\mathtt{eff}}=\frac{\varepsilon +mc^{2}}{%
mc^{2}}\,V_{s}  \label{9}
\end{equation}

\noindent the KG equation can be written as

\begin{equation}
-\frac{\hbar ^{2}}{2m}\,\phi ^{\prime \prime }+V_{\mathtt{eff}}\,\phi =E_{%
\mathtt{eff}}\,\phi  \label{11}
\end{equation}%
On the other hand, for $\varepsilon \neq -mc^{2}$ the same Sturm-Liouville
equation for $\phi $ is obeyed by $\psi _{+}$ whereas $\psi _{-}=-i\hbar
c\psi _{+}^{\prime }/\left( \varepsilon +mc^{2}\right) .$ Otherwise, for $%
\varepsilon =-mc^{2}$, it might be possible the existence of an isolated
solution given by%
\begin{equation}
\psi _{+}=\mathrm{const.},\quad \psi _{-}=\frac{2\psi _{+}}{i\hbar c}%
\int^{x}dx\left( V_{s}+mc^{2}\right)  \label{14}
\end{equation}%
Of course, this solution does not exist if the domain is infinity because $%
\psi _{+}$ would not be square integrable. Note that apart from the possible
isolated solution, $\psi _{+}$ satisfies the KG equation. An equally
interesting result in the case of vanishing mass is that the spectrum just
changes sign when $V_{0}$ does. As for the eigenfunctions, $\phi $ and $\psi
_{+}$ are invariant under the change of the sign of $V_{0}$ whereas $\psi
_{-}$ changes sign.

Let us consider the specific case of the two-parameter potential functions $%
V_{v}=V_{0}\,\mathrm{sec}^{2}\,\alpha x$ and $V_{s}=V_{0}\,\mathrm{tan}%
^{2}\,\alpha x$. In this case the isolated solution of the Dirac equation
for $\psi _{-}$ is not normalizable and the effective potential of the
Sturm-Liouville problem for both $\phi $ and $\psi _{+}$ can be expressed as%
\begin{equation}
V_{\mathtt{eff}}=U_{0}\,\mathrm{tan}^{2}\,\alpha x,\quad U_{0}=\frac{%
\varepsilon +mc^{2}}{mc^{2}}V_{0}  \label{15}
\end{equation}%
Notice that $V_{\mathtt{eff}}$ is invariant under the change $\alpha
\rightarrow -\alpha $ so that the results can depend only on $|\alpha |$.
Furthermore, the effective potential is an even function under $x\rightarrow
-x$ in such way that $\phi $ and $\psi _{+}$ can be taken to be even or odd.
When $\varepsilon <-mc^{2}$ for $V_{0}>0$ and $\varepsilon >-mc^{2}$ for $%
V_{0}<0$ one has $U_{0}<0$. In this case the effective potential consists of
periodical wells and barriers. On the other hand, \ when $\varepsilon
>-mc^{2}$ for $V_{0}>0$ and $\varepsilon <-mc^{2}$ for $V_{0}<0$ one has $%
U_{0}>0$ and the effective potential is identified as the exactly solvable
symmetric P\"{o}schl-Teller potential \cite{rm}-\cite{nie1}. In this last
circumstance, due to the infinities at $|x|=\pi /(2|\alpha |)$, attention
can be restricted to $|x|<\pi /(2|\alpha |)$. In fact, the effective
potential is a well potential limited by infinite barriers at $x=\pm \pi
/(2|\alpha |)$ so that the capacity of the effective potential to hold
bound-state solutions with $\varepsilon >mc^{2}$ for $V_{0}>0$ and $%
\varepsilon <-mc^{2}$ for $V_{0}<0$ is infinite (with a spectral gap in the
interval $|\varepsilon |<mc^{2}$ for $V_{0}>0$).

For the bound-state solutions, one can see that the normalizable
eigenfunctions are subject to the boundary conditions $\phi =\psi _{+}=0$ as
$|x|=\pi /(2|\alpha |)$ (where the potential becomes infinitely steep) in
such a manner that the solution of our relativistic problem can be developed
by taking advantage from the knowledge of the exact solution for the
symmetric P\"{o}schl-Teller potential. The corresponding effective
eigenenergy is given by \cite{rm}-\cite{nie1}
\begin{equation}
\frac{\varepsilon ^{2}-m^{2}c^{4}}{2mc^{2}}=\,\frac{\hbar ^{2}\alpha ^{2}}{2m%
}\left( n^{2}+2n\lambda +\lambda \right) ,\quad n=0,1,2,\ldots ,\quad
\label{16}
\end{equation}

\noindent where
\begin{equation}
\lambda =\frac{1}{2}\left( 1+\sqrt{1+\frac{8mU_{0}}{\hbar ^{2}\alpha ^{2}}}%
\,\right)  \label{17}
\end{equation}%
\noindent Now, (\ref{16})-(\ref{17}) lead to the quantization condition%
\begin{equation}
2\sqrt{\varepsilon ^{2}-m^{2}c^{4}+2\left( \varepsilon +mc^{2}\right) V_{0}}-%
\sqrt{\hbar ^{2}c^{2}\alpha ^{2}+8V_{0}\left( \varepsilon +mc^{2}\right) }%
=\hbar c|\alpha |\left( 2n+1\right)  \label{19}
\end{equation}%
The solutions of (\ref{19}) determinate the eigenvalues of the relativistic
problem. This equation can be solved easily with a symbolic algebra program
by searching eigenenergies in the range $\varepsilon >mc^{2}$ for $V_{0}>0$
and $\varepsilon <-mc^{2}$ for $V_{0}<0$, as foreseen by the preceding
qualitative arguments. Of course, for $V_{0}>0$ one obtains $\varepsilon
\approx mc^{2}$ for the lowest quantum numbers when $V_{0}\ll mc^{2}$. One
the other hand, for $V_{0}<0$ one finds $\varepsilon \approx -mc^{2}$ for
the lowest quantum numbers when $|V_{0}|\ll mc^{2}$. It happens that there
is at most one solution of (\ref{19}) for a given quantum number. Figures 1
and 2 show the behaviour of the energies as a function of \ $V_{0}$ and $%
\alpha $, respectively. It is noticeable from both of these figures, for $%
V_{0}>0$, that for a given set of potential parameters one finds that the
lowest quantum numbers correspond to the lowest eigenenergies, as it should
be for particle energy levels. For $V_{0}<0$ the spectrum presents a similar
behavior but the the highest energy levels are labelled by the lowest
quantum numbers and are to be identified with antiparticle levels. If we had
plotted the spectra for a massless particle, we would encounter, up to the
sign of $\varepsilon $, identical spectra for both signs of $V_{0}$. At any
circumstance, the spectrum contains either particle-energy levels or
antiparticle-energy levels.

The KG eigenfunction as well as the upper component of the Dirac spinor can
be given by \cite{nie1}
\begin{equation}
\phi =\psi _{+}=N\,2^{-\lambda }\sqrt{2|\alpha |\left( n+\lambda \right)
\frac{\Gamma \left( n+1\right) }{\Gamma \left( n+2\lambda \right) }}\frac{%
\Gamma \left( 2\lambda \right) }{\Gamma \left( \lambda +\frac{1}{2}\right) }%
\left( 1-z^{2}\right) ^{\lambda /2}C_{n}^{\left( \lambda \right) }\left(
z\right)  \label{21}
\end{equation}

\noindent where $z=\mathrm{sin}\,\alpha x$ and $C_{n}^{\left( \lambda
\right) }\left( z\right) $ is the Gegenbauer (ultraspherical) polynomial of
degree $n$. Since $C_{n}^{\left( \lambda \right) }\left( -z\right) =\left(
-\right) ^{n}C_{n}^{\left( \lambda \right) }\left( z\right) $ and $%
C_{n}^{\left( \lambda \right) }\left( z\right) $ has $n$ distinct zeros
(see, e.g. \cite{abr}), it becomes clear that $\psi _{+}$ and $\psi _{-}$
have definite and opposite parities. The constant $N$ \ is the unit in the
KG problem and it chosen such that $\int_{-\infty }^{+\infty }dx\left( |\psi
_{+}|^{2}+|\psi _{-}|^{2}\right) =1$ in the Dirac problem. Fig. 3
illustrates the behaviour of the upper and lower components of the Dirac
spinor $|\psi _{+}|^{2}$ and $|\psi _{-}|^{2}$, and the position probability
densities $|\psi |^{2}=|\psi _{+}|^{2}+|\psi _{-}|^{2}$ and $|\phi |^{2}$
for $n=0$. The relative normalization constant was calculated numerically.
Comparison of $|\psi _{+}|^{2}$ and $|\psi _{-}|^{2}$ shows that $\psi _{-}$
$\ $\ is suppressed relative to $\psi _{+}$. This result is expected since
we have here an particle eigenstate. Surprisingly, the same behaviour shows
its face for the antiparticle eigenstates (for $V_{0}<0$). In addition,
comparison of $\ |\phi |^{2}|$ and $|\psi |^{2}$ shows that a KG particle
tends to be better localized than a Dirac particle.

In summary, the methodology for finding solutions of the KG and the Dirac
equations for the enlarged class of mixed vector-scalar potentials
satisfying the constraint $V_{v}=V_{s}+V_{0}$ have been put forward. With
the two-parameter potential functions $V_{v}=V_{0}\,\mathrm{sec}^{2}\,\alpha
x$ and $V_{s}=V_{0}\,\mathrm{tan}^{2}\,\alpha x$, the KG equation and the
Dirac equation for $\psi _{+}$ have been mapped into a Schr\"{o}dinger-like
equation with the symmetric P\"{o}schl-Teller potential. The spectrum of
these relativistic problems consists of infinitely many discrete
eigenenergies related to either particle or antiparticle levels in such a
way the Klein%
\'{}%
s paradox is absent from the scenario. As has been commented above, changing
the sign of $V_{v}$ allows us to migrate from the particle sector to the
antiparticle sector and vice versa just by changing the sign of the
eigenenergies as far as the spectra is concerned. These changes imply that  $%
|\phi |$ maintains its nodal structure whereas $|\psi _{+}|$ and $|\psi _{-}|
$ exchange theirs in such a way that the nodal structure of the position
probability density is preserved. Although the KG and the Dirac equations
exhibit the very same spectrum their eigenfunctions make all the difference.
In fact, we have shown that a KG particle tends to be better localized than
a Dirac particle.

\bigskip \bigskip

\noindent \textbf{Acknowledgments}

This work was supported in part by means of funds provided by CAPES, CNPq
and FAPESP.

\newpage

\begin{figure}[th]
\begin{center}
\includegraphics[width=9cm, angle=270]{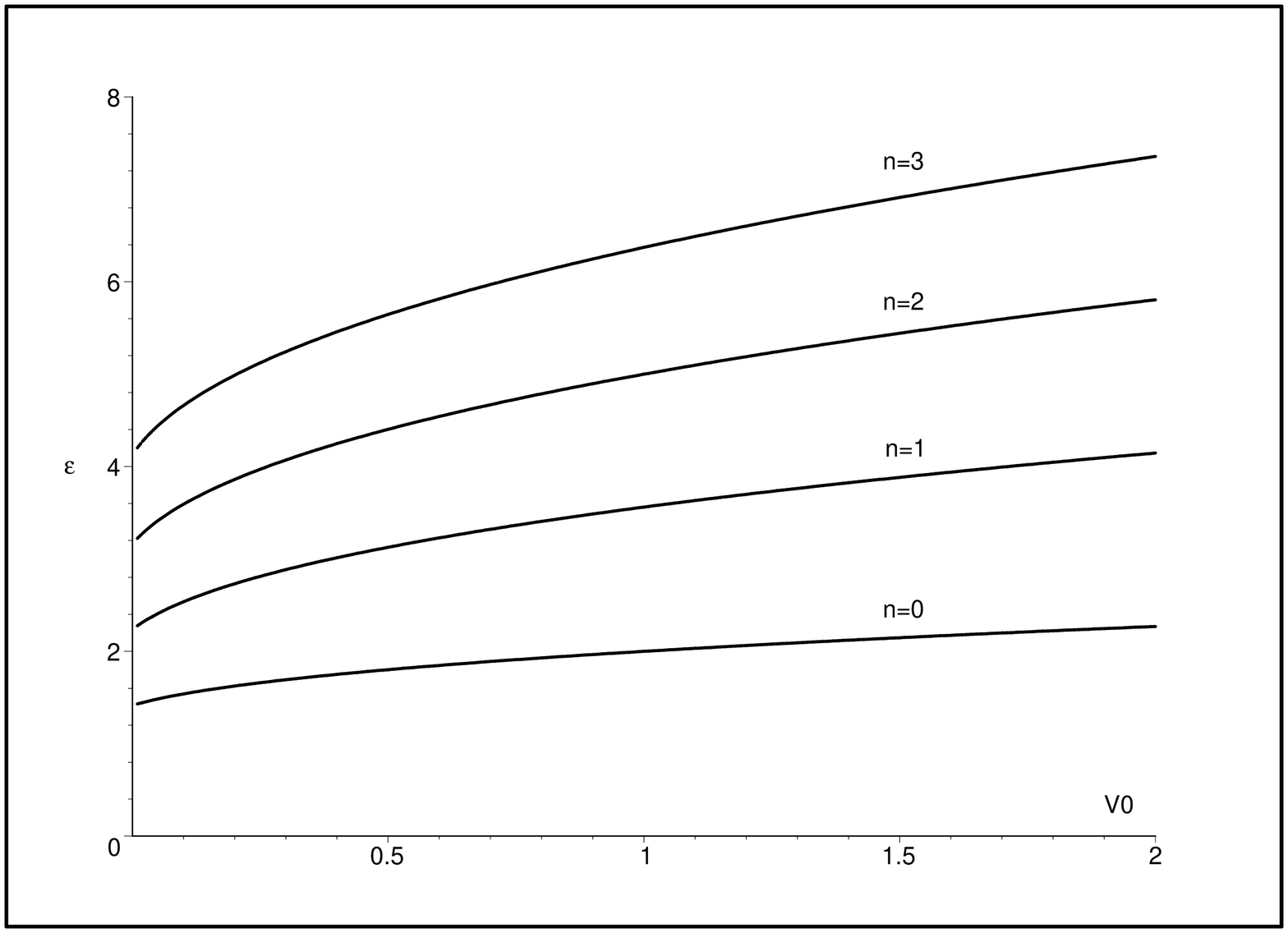}
\end{center}
\par
\vspace*{-0.1cm}
\caption{Dirac eigenvalues, in the sense of $\protect\varepsilon$, for the
four lowest quantum numbers as a function of $V_{0}$ ($m=\hbar=c=\protect%
\alpha = 1$). }
\label{Fig1}
\end{figure}

\begin{figure}[th]
\begin{center}
\includegraphics[width=9cm, angle=270]{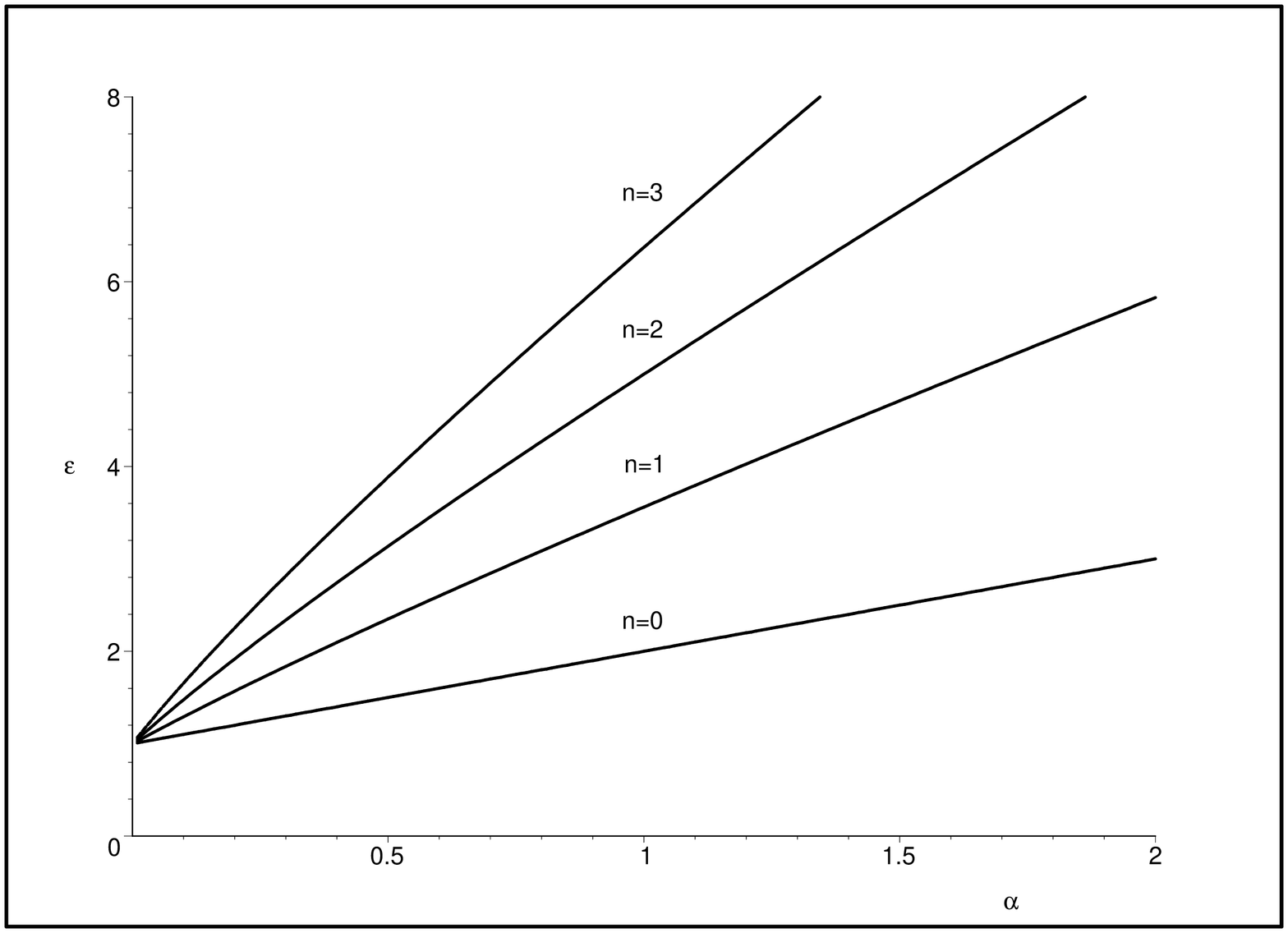}
\end{center}
\par
\vspace*{-0.1cm}
\caption{Dirac eigenvalues, in the sense of $\protect\varepsilon$, for the
four lowest quantum numbers as a function of $\protect\alpha$ ($m=\hbar=c=1$
and $V_{0} = 1$). }
\label{Fig2}
\end{figure}

\begin{figure}[th]
\begin{center}
\includegraphics[width=9cm, angle=270]{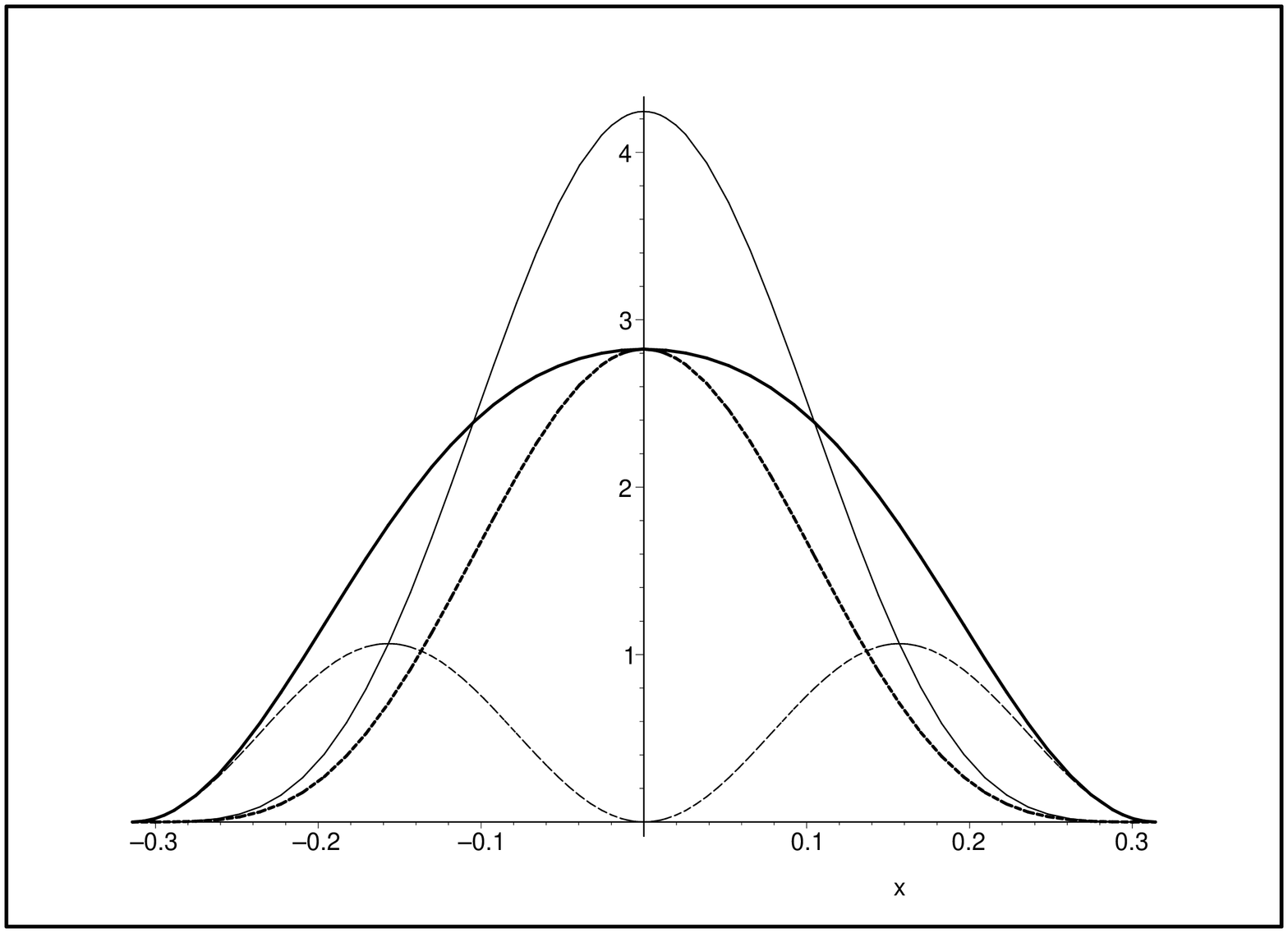}
\end{center}
\par
\vspace*{-0.1cm}
\caption{$|\protect\psi _{+}|^{2}$ (heavy dashed line), $|\protect\psi %
_{-}|^{2}$ (light dashed line), $|\protect\psi |^{2}=|\protect\psi %
_{+}|^{2}+|\protect\psi _{-}|^{2}$ (thick line) and $|\protect\phi |^{2}$
(thin line) for $n=0$ ($m=\hbar=c=1$, $V_{0}=3$ and $\protect\alpha=5$). }
\label{Fig3}
\end{figure}


\newpage

\begin{thebibliography}{9}
\bibitem{luc} W. Lucha et al., Phys. Rep. 200 (1991) 127 and references
therein.

\bibitem{ser} B.D. Serot, J.D. Walecka, in: Advances in Nuclear Physics,
Vol. 16, edited by J.W. Negele and E. Vogt, Plenum, New York, 1986; J.N.
Ginocchio, Phys. Rev. Lett. 78 (1997) 436; J.N. Ginocchio, A. Leviatan,
Phys. Lett. B 425 (1998) 1; J.N. Ginocchio, Phys. Rep. 315 (1999) 231; P.
Alberto et al., Phys. Rev. Lett. 86 (2001) 5015; P. Alberto et al., Phys.
Rev. C 65 (2002) 034307; T.-S. Chen et al., Chin. Phys. Lett. 20 (2003) 358;
G. Mao, Phys. Rev. C 67 (2003) 044318; R. Lisboa et al., Phys. Rev. C 69
(2004) 024319.

\bibitem{gum} G. Gumbs, D. Kiang, Am. J. Phys. 54 (1986) 462; F. Dom\'{\i}%
nguez-Adame, Am. J. Phys. 58 (1990) 886; A.S. de Castro, Phys. Lett. A 305
(2002) 100; Y. Nogami et al., Am. J. Phys. 71 (2003) 950; A.S. de Castro,
Phys. Lett. A 338 (2005) 81; A.S. de Castro, Phys. Lett. A 346 (2005) 71;
A.S. de Castro, Ann. Phys. (N.Y.) 316 (2005) 414; A.S. de Castro,
hep-th/0507025, Int. J. Mod. Phys. A, in press; A.S. de Castro,
hep-th/0511010, Int. J. Mod. Phys. A, in press; A.S. de Castro et al., Phys.
Rev. C 73 (2006) 054309.

\bibitem{rm} G. P\"{o}schl, E. Teller, Z. Phys. 83 (1933) 143; I.I. Gol\'{}%
dman, V.D. Krivchenkov, Problems in Quantum Mechanics, Pergamon, London,
1961.

\bibitem{nie1} M.M. Nieto, L.M. Simmons Jr., Phys. Rev. Lett. 41 (1978) 207;
M.M. Nieto, L.M. Simmons Jr., Phys. Rev. D 20 (1979) 1332; M.M. Nieto, Phys.
Rev. A 20 (1979) 700.

\bibitem{bra} See, e.g., O.M. Braun, Y.S. Kivshar, The Frenkel-Kontorova
Model: Concepts, Methods, and Applications, Springer, Berlin, 2004.

\bibitem{abr} M. Abramowitz, I.A. Stegun, Handbook of Mathematical
Functions, Dover, Toronto, 1965.

\bibitem{str} W. Greiner, Relativistic Quantum Mechanics: Wave Equations,
Springer-Verlag, Berlin, 1990; P. Strange, Relativistic Quantum Mechanics,
Cambridge University Press, Cambridge, 1998. '
\end{thebibliography}
\end{document}